\documentclass[
aps,
prd,
reprint,
longbibliography, 
amsmath, 
amssymb, 
nofootinbib,
superscriptaddress
]{revtex4-2}

\usepackage[colorlinks=true, allcolors=black]{hyperref}
\usepackage{enumitem}
\usepackage[normalem]{ulem}
\usepackage{comment}
\usepackage[super]{nth}
\usepackage{graphicx}
\usepackage{dcolumn}
\usepackage{bm}
\usepackage[dvipsnames]{xcolor}
\usepackage{pgfplotstable, booktabs}
\usepackage{array}

\usepackage{amsmath}
\usepackage{xcolor}

\usepackage{float}
\usepackage{enumitem}

\newcommand\einstein{\texttt{Einstein Toolkit}}
\newcommand\ilgrmhd{\texttt{IllinoisGRMHD}}

\renewcommand{\d}[1]{\ensuremath{\operatorname{d}\!{#1}}}

\UseRawInputEncoding

\begin{document}

\title{Dual Horizon-MAD: A New Mode of Binary Black Hole Magnetically Arrested Accretion with Saturated Horizon Magnetic Fluxes and Intermittent Minidisks}

\author{Vikram Manikantan}
\email{vik@arizona.edu}
\affiliation{Steward Observatory \& Department of Astronomy, University of Arizona, Tucson, Arizona 85721, USA}

\author{Vasileios Paschalidis}
\email{vpaschal@arizona.edu}
\affiliation{Steward Observatory \& Department of Astronomy, University of Arizona, Tucson, Arizona 85721, USA}
\affiliation{Department of Physics, University of Arizona, Tucson, Arizona 85721, USA}
\date{\today}

\begin{abstract}
Reliable first-principles predictions of multimessenger signals from inspiralling supermassive binary black holes (SMBBHs) require magnetohydrodynamic (MHD) simulations in full general relativity (GR) that resolve the accretion flow down to the BH horizons. However, performing such simulations long enough to relax the magnetohydrodynamics near the BBHs is computationally challenging and a source of uncertainty in modeling efforts. In this work, we present the longest MHD simulation of circumbinary disk (CBD) accretion onto a SMBBH in full GR. The equal-mass, quasicircular, nonspinning BBH starts at an initial separation of $d = 30 \, GM/c^2$ and evolves over $\sim56$ orbits ($t \simeq 52,000 \, GM/c^3$), relaxing the magnetic field dynamics within the CBD cavity. After $\sim 23$ orbits and persistent magnetically arrested minidisks, the accretion enters a new magnetically arrested mode which we call dual horizon-MAD (DH-MAD). In DH-MAD, the dimensionless horizon magnetic fluxes remain saturated at $\phi_{BH}\sim 20$ driving magnetic flux eruptions from the vicinity of the BH horizons, and the vertical magnetic flux in the cavity saturates, but no flux eruptions occur at the cavity-CBD boundary. During this DH-MAD phase, there is also a shift in the total mass accretion rate variability, which exhibits a periodicity of $f\sim 1.8 f_{\rm orb}$, while the accretion rate modulation on the individual black holes occurs almost at the orbital period, $f\sim 0.9 f_{\rm orb}$. $\phi_{BH}$ also exhibits a robust periodicity at $f\sim 0.4 f_{\rm orb}$, suggesting that quasiperiodic nonthermal emission from magnetic reconnection near the horizons could be a smoking gun electromagnetic counterpart to inspiraling SMBBHs. Finally, minidisks around the BHs become intermittent, even at a separation of $d = 24 \, GM/c^2$, likely due to the large plasma magnetization in the optically thin cavity. DH-MAD accretion may produce distinctive quasi-periodic electromagnetic signatures in SMBBHs, facilitating the identification of electromagnetic counterparts to low-frequency gravitational wave sources.
\end{abstract}

\maketitle

\section{Introduction}\label{sec:intro}

Supermassive binary black holes (SMBBHs) are expected to form ubiquitously in gas-rich environments as the product of galaxy mergers~\cite{begelman_massive_1980}. Magnetized gas accretion onto the SMBBH can power relativistic dual-jets~\cite{farris_binary_2012,gold_accretion_2014, Gold:2013zma, de_simone_kicked_2026, ennoggi_merger_2025}, magnetic flux eruptions~\cite{manikantan_magnetically_2025, most_magnetically_2024, wang_bmad_2025, ressler_dual_2025}, and a luminous accretion flow, driving thermal and nonthermal emission across the electromagnetic (EM) spectrum~\cite{bogdanovic_electromagnetic_2022}. Additionally, during their late inspiral and merger, SMBBHs emit gravitational waves (GWs) detectable by the Laser Interferometer Space Antenna (LISA)~\cite{LISA_white_paper} and Pulsar Timing Arrays (PTA)~\cite{nanograv}. This makes SMBBHs among the most promising multimessenger sources with GWs in our universe. Coincident multimessenger observations of SMBBHs are expected to help answer fundamental questions in astronomy, astrophysics, cosmology, and fundamental physics~\cite{lisa_multimessenger_whitepaper, lisa_gw_whitepaper}.

Once a typical $10^6-10^7M_\odot$ SMBBH is within a few milliparsecs, GW emission will drive the binary to merger within a Hubble time, where the inspiral time is given by $t_{\rm gw}\sim  2.3\times 10^{9} (d/1 {\rm mpc} )^4 (M_{\rm BBH}/10^6 M_\odot)^{-3} \rm yrs$~\cite{peters_gravitational_1964, shapiro_black_1983}. When the binary separation reaches $d\sim 30 \, GM/c^2$ ($\sim  10 \mu \rm pc$ for $M_{\rm BBH} = 10^7 M_{\rm \odot}$), the SMBBH 
enters the late inspiral, relativistic phase, and
is about $100/\tilde\eta$ orbits ($\tilde\eta\equiv 4\eta$ with $\eta= q/(1+q)^2$ being the symmetric mass ratio and $q=m_1/m_2$) from merger. This is the regime where LISA and PTAs will detect individual SMBBHs, and where we will be able to perform ground-breaking multimessenger science with SMBBHs~\cite{lisa_multimessenger_whitepaper}.

However, detecting a SMBBH is not trivial. PTAs, which probe nanohertz frequencies, have measured a stochastic background GW emission that could be explained by a population of inspiraling SMBBHs~\cite{agazie_nanograv_2023, nanograv}, but an individual detection is still elusive~\cite{rosado_expected_2015, kelley_single_2018, aggarwal_nanograv_2019}. LISA will detect millihertz GWs from $M_{\rm BBH} \sim 10^{4-7} M_{\odot}$ inspiraling SMBBHs (among other sources) but its launch is scheduled for 2035~\cite{LISA_white_paper, lisa_gw_whitepaper, lisa_paper, amaro-seoane_astrophysics_2023, lisa_consortium_waveform_working_group_waveform_2023, arun_new_2022}. Even with a GW detection, performing multimessenger studies of SMBBHs faces a significant obstacle: the sky localization of GW sources is poor. For example, the sky localization with LISA can range from $1\text{ to }100 \rm \, deg^2$ depending on the distance and orientation of the source with respect to the detector~\cite{mangiagli_observing_2020, piro_chasing_2023}. Therefore, identifying a host galaxy without a robust theoretical understanding of EM counterparts to the inspiraling SMBBH will be challenging~\cite{lops_galaxy_2023, drake_preliminary_2025}. This makes understanding the EM emission from SMBBHs, and how they distinguish themselves from single SMBHs, an urgent task for multimessenger astronomy~\cite{decadal_survey_on_astronomy_and_astrophysics_2020_astro2020_pathways_2021}. 

EM surveys have revealed $100$s of SMBBH candidates, including dozens in the late inspiral regime~\cite{rodriguez_2006, rodriguez_2009, charisi_multiple_2015, graham_systematic_2015, panstars_smbh, oneill_unanticipated_2022, bright_minidisk_2023, kiehlmann_pks_2024}. However, without a direct GW detection from these sources or a horizon-scale observation (with, e.g., the Event Horizon Telescope; EHT~\cite{eht_m87, eht_saga_2022}) we cannot confirm these candidates as SMBBHs. Therefore, to vet existing and future candidates as true SMBBH sources, as well as to identify the host galaxies of future GW sources, we need reliable models of SMBBH lightcurves and spectra derived from high-fidelity numerical simulations from first principles. 

To understand long-term processes that occur on hundreds of orbits at large orbital separations, such as angular momentum transfer between the binary and gas, studies use Newtonian simulations of hydrodynamic accretion (see e.g., ~\cite{westernacher-schneider_eccentric_2024, westernacher-schneider_multi-band_2022,Garg:2026xgp}). Most Newtonian studies model 2D thin disks, and hence are restricted to accretion rates $\dot{M} \sim 10^{-2} \dot{M}_{\rm edd}$, where $\dot{M}_{\rm edd}$ is the Eddington accretion rate. 2D studies neglect the evolution of magnetic fields (but see \cite{avara_magnetic_2026} for a recent attempt to model magnetic fields in 2D) and do not account for jet feedback. To perform hundreds to a thousand orbits in three dimensions, Newtonian studies sometimes excise the region within the binary orbit or within the CBD cavity disk~\cite{Shilump2012ApJ...749..118S,Garg:2026xgp}. Such inner boundary conditions miss physics that is important for angular momentum transport, and accretion onto the individual BHs, see, e.g.,~\cite{Roedig2012A&A...545A.127R}. Newtonian 3D MHD studies have revealed that large-scale, vertical magnetic fields can saturate the low-density cavity region within a circumbinary disk (CBD) and perturb accretion onto the binary~\cite{most_magnetically_2024, most_decoupling_2025, wang_bmad_2025}. While Newtonian studies provide useful insight because they can perform simulations long enough to relax the CBD, they cannot model feedback from the BH horizons self-consistently. The artificial inner boundary conditions they impose can qualitatively and quantitatively alter the accretion flow, especially near the BHs. This can potentially affect the cavity and predictions of the EM signal that emerges from that region. Therefore, to provide high-fidelity EM models of SMBBHs during their late inspiral and merger, we must use numerical simulations that simultaneously solve Einstein's equations of general relativity (GR) and the equations of magnetohydrodynamics (MHD). 

Previous studies of CBD MHD accretion onto BBHs in full GR have revealed a few key features: 1. accretion onto the BBH is quasi-periodic~\cite{farris_binary_2012, farris_binary_2014, khan_disks_2018, cattorini_misaligned_2022, cattorini_grmhd_2024, ruiz_general_2023, manikantan_effects_2025, manikantan_coincident_2025, ennoggi_merger_2025}, 2. the BBH launches dual-jets, even for nonspinning BBHs~\cite{farris_binary_2012,Gold:2013zma,Gold:2013zma,paschalidis_minidisk_2021, bright_minidisk_2023, ennoggi_relativistic_2025, de_simone_kicked_2026}, 3. the BHs can host persistent minidisks if the ratio between the Hill sphere radius and innermost stable circular orbit (ISCO) is large enough~\cite{paschalidis_minidisk_2021, bright_minidisk_2023}, and under certain conditions the BHs can host magnetically arrested minidisks (MAMs; \cite{manikantan_magnetically_2025}). Studies in approximate spacetimes also demonstrate similar results see, e.g., ~\cite{noble_circumbinary_2012,avara_accretion_2024, combi_minidisk_2022, combi_magnetized_2026, avara_accretion_2024, kim_effects_2026, zhu_finite-solid-angle_2026}. Relaxing the entire disk in full GR is extremely challenging. Efforts have been made to relax the fluid prior to taking over with a full GR simulation, but often do not include or underresolve the BH horizons and the cavity in the evolution~\cite{noble_circumbinary_2012, avara_accretion_2024, ennoggi_merger_2025}. This is the focus of the current work: relaxing the fluid and magnetic field behavior within the cavity while evolving the BBH spacetime in full GR.

In this work, we present the longest to-date MHD  simulation of CBD accretion onto an inspiraling SMBBH in full GR. The binary initial coordinate separation is $d= 30 \, r_g$, ($r_g=GM/c^2$ with $M$ the binary gravitational mass), and evolves down to $d \sim 24 \, r_g$ ($t \sim 52,000 r_g/c$; $N_{\rm orbits}\sim 56$). The long integration time allows for the development of new steady state mode of accretion that we call dual horizon magnetically arrested disk (DH-MAD). In this accretion mode, the dimensionless horizon magnetic fluxes $\phi_{\rm BH}$ remain saturated at $\phi_{BH}\sim 20$ and drive magnetic flares as in MADs~\cite{narayan_mad_2003, igumenshchev_mad_2003, tchekhovskoy_mad_2011, mckinney_mad_2012}.  The vertical magnetic flux in the cavity reaches saturation, but no eruptions from the cavity-disk boundary occur as in BMAD flows~\cite{most_magnetically_2024, wang_bmad_2025}. DH-MAD is a distinct mode because the magnetic flares are driven only by the magnetic flux saturation on the two BH horizons rather than the cavity. In the DH-MAD phase there is also a shift in the total mass accretion rate variability, which exhibits a periodicity of $f\sim 1.8 f_{\rm orb}$, while the accretion rate modulation on the individual black holes occurs almost at the orbital period, at $f\sim 0.9 f_{\rm orb}$. We also find that in the DH-MAD phase flux eruptions from the horizon vicinity continue quasiperiodically, current sheets form near the horizon and through the cavity, and that the toroidal magnetic field dominates over the poloidal magnetic field in the bulk of the CBD cavity.  Finally, minidisks around the BHs become intermittent, despite being at a large separation of $d = 24 \, r_g$, where the Hill radius comfortably accommodates the innermost stable orbit of each BH.

The paper is structured as follows. In Section~\ref{sec:methods} we briefly discuss the initial data, evolution methods, and detail the diagnostic tools we adopt. In Section~\ref{sec:results} we outline the key features of the CBD cavity and  accretion flow. Finally, in Section~\ref{sec:summary} we summarize and discuss our findings. Throughout, we adopt geometrized units in which $G=c=1$, where $G$ is the gravitational constant and $c$ is the speed of light. Our spatial and time domains are measured in units of $M$, where $M$ is the Arnowitt-Deser-Misner (ADM) mass of the spacetime. Therefore, in the spatial domain $1 \,M = 1 \, r_g$ and in the time domain $1M = 1 \, r_g/c \equiv GM/c^3$. 

\begin{figure*}
    \centering
    \includegraphics[width=1\linewidth]{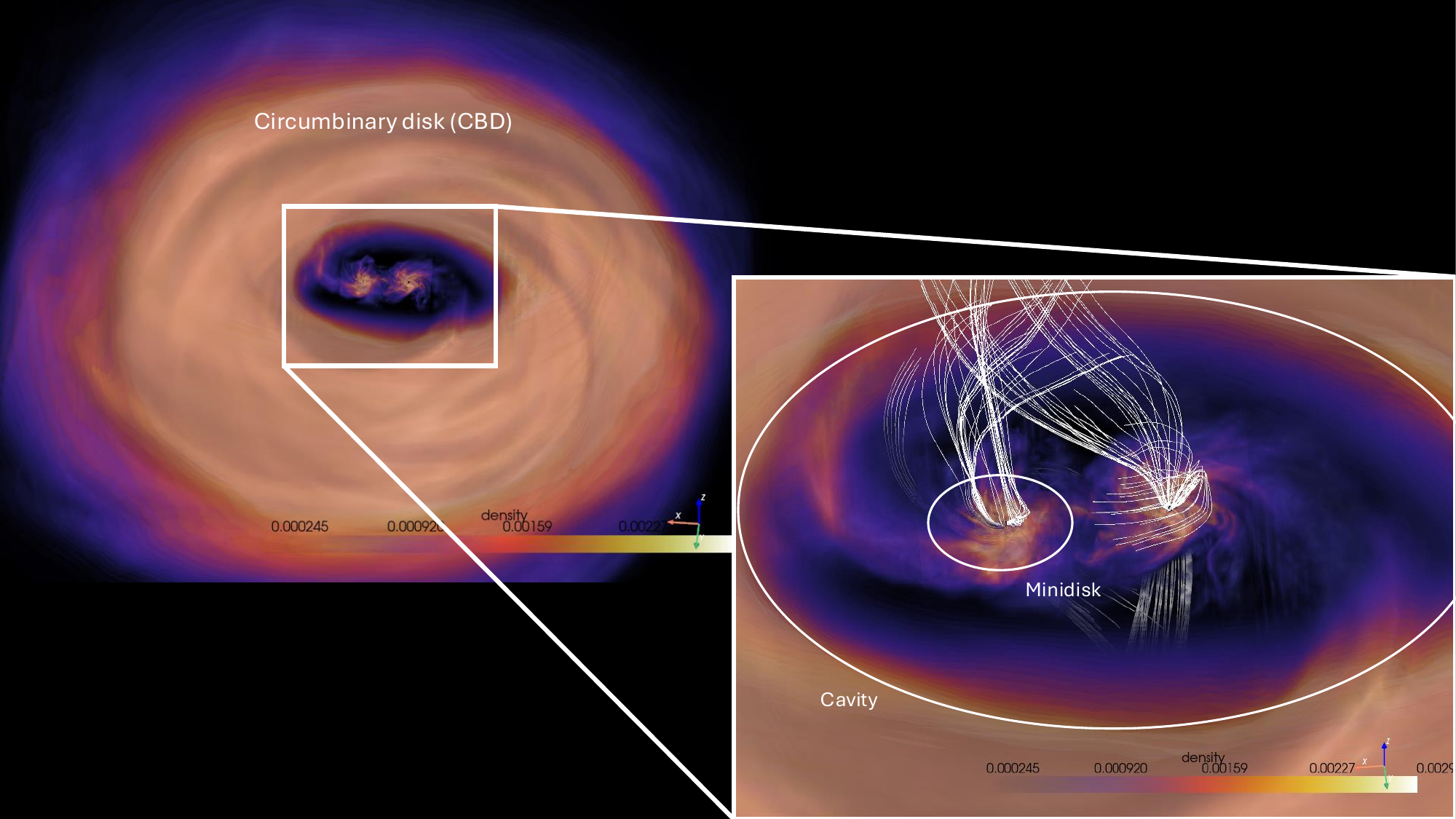}
    \caption{Volume rendering of the rest-mass gas density in our MHD simulation in full GR of a circumbinary disk (CBD) accreting onto an equal-mass, nonspinning BBH at $t/M = 46,339.2$. The BHs have inspiraled from an initial separation of $d/M = 30$ to $d/M \simeq 24.7$. In the top left panel, we show the entire CBD with density contours of minimum $\rho_0/\rho_{0, \rm max} \sim 2 \times 10^{-4}$ and maximum $\rho_0/\rho_{0, \rm max} \sim 3 \times 10^{-3}$. We indicate the cavity region and one of the two minidisks with white circles and labels. We plot the magnetic fields threading the BH horizons with white streamlines. The magnetic field lines are not significantly twisted as the BHs are nonspinning.}
    \label{fig:3d_plot}
\end{figure*}

\section{Methods} \label{sec:methods}
\subsection{Simulation Set-up}
    The details of our simulation set up and evolution have been outlined in~\cite{manikantan_magnetically_2025}. To briefly summarize, we perform the simulation within the \einstein{} framework~\cite{einsteintoolkit}. We use the \texttt{TwoPunctures} code to prepare the spacetime initial data~\cite{twopunctures_ansorg, twopunctures_paschalidis} such that the BHs have no spin ($\chi_1 = \chi_2 = 0$), are equal mass ($q = m_2/m_1 = 1$), and are on a quasicircular orbit with separation $d/M = 30$. The BH horizons are resolved with $\sim 64$ cells, with the resolution at the highest level of $\Delta x = M/64$ and outer boundary of the grid at $r_{\rm out}/M = 5120$. We seed our powerlaw CBD with a poloidal magnetic field by setting the vector potential as,
    
    \begin{equation}
        (A_x, A_y) = (-y, x) (x^2 + y^2)^{-1/2} A_b \, {\rm max}(P - 0.1 P_{\rm max}, 0),
    \end{equation}
    where $A_b=15$~\cite{chakrabarti_1985, khan_disks_2018}. This results in an initial minimum plasma beta $\beta \equiv P_{\rm gas}/P_{\rm mag} \simeq 26$. The initial plasma beta parameter as defined by $\beta_{\rm ini} \equiv 2\, \max(P_{\rm gas})/\max(b^2)$ is $\beta_{\rm ini} = 52.42$. 

    We evolve the spacetime by solving the Einstein equations in the Baumgarte-Shapiro-Shibata-Nakamura (BSSN) formalism~\cite{Shibata1995, Baumgarte1998}, as implemented in the \texttt{LeanBSSN} thorn using 6th-order finite differences~\cite{sperhake_binary_2007}.  We employ the $3+1$ general-relativistic MHD AMR code \ilgrmhd{} to evolve the fluid and magnetic fields~\cite{ilgrmhd}.

    \subsection{Diagnostics}
    \subsubsection{Horizon Magnetic Flux}
    \label{subsec:b_flux}

    The horizon magnetic flux $\Phi_{\rm BH}$ is computed as described in the Appendix of~\cite{manikantan_magnetically_2025}, and implemented in the publicly available {\texttt{QuasiLocalMeasuresMHD}}\footnote{https://github.com/arizona-relativity/QuasiLocalMeasuresMHD} thorn. The dimensionless magnetic flux is given by
    \begin{equation}
    \label{eq:phibh}
        \phi_{\rm BH} = \Phi_{\rm BH} / \sqrt{\langle \dot{M}_{\rm BH} \rangle r_{\rm g}^2 c},
    \end{equation}
    where $\langle \dot{M}_{\rm BH} \rangle$ is the running average of the rest-mass accretion rate onto the respective BH horizon, $r_g \equiv GM_{\rm BH}/c^2$ with $M_{\rm BH}$ being the mass of the individual BH, and $c=1$ the speed of light in our calculations. Note that in {\tt IllinoisGRMHD}~\cite{ilgrmhd} the magnetic field measured by a normal observer is already in Gaussian units; no factors of $\sqrt{4\pi}$ are absorbed in the Faraday tensor by contrast to the original {\tt HARM} formulation~\cite{Gammie:2003rj} which has formed the basis of many GRMHD accretion codes. As such our flux normalization agrees with the standard convention adopted in MAD studies~\cite{narayan_jets_2022}. We locate BH apparent horizons, on which $\Phi_{\rm BH}$ is measured, with the \texttt{AHFinderDirect} thorn~\cite{Thornburg2004}
    
    \subsubsection{Cavity Magnetic Flux}
    We also report the cavity magnetic flux by integrating the vertical magnetic flux through disk-like surfaces of varying coordinate radius at $z=0$. We exclude the BH horizons in the integration. We define the cavity flux as 

    \begin{equation}
        \begin{split}
        \Phi_c(r) &= \int_{S} (1/2)F_{\mu\nu} \d x^{\mu}\wedge \d x^\nu \\
        &= \int_{S} \sqrt{\gamma} B^z \, \d x \d y
        \end{split}
    \end{equation}
where $S$ is the surface of a disk of coordinate radius $r$ on the orbital ($xy$) plane, $F_{\mu\nu}$ is the antisymmetric Faraday tensor, and $B^z$ the z-component of the magnetic field measured by a normal observer (see Appendix of~\cite{manikantan_magnetically_2025} for a more detailed derivation of the flux integral). The integration domain is centered on the binary center-of-mass and the disk radius $r$ is set to fractions of the cavity coordinate radius $r_c = d$, with $d$ the time-dependent BBH separation. 

    The dimensionless fluxes are given by
    \begin{equation}
    \label{eq:cavity_flux}
        \phi_c = \Phi_c(r) / \sqrt{\langle \dot{M}_{\rm tot} \rangle r_{\rm g}^2 c},
    \end{equation}
    where $\langle \dot{M}_{\rm tot} \rangle$ is the running average of the total mass accretion rate onto the BH horizons.

     We normalize the flux value to the total accretion rate onto the BHs, which we expect to be close to the mass flux into the cavity as our system has reached steady-state within the innermost parts of the disk, as indicated by the settled mass accretion rate onto the horizons. However, reaching a true radial inflow steady state throughout the entire disk may take more than $t/M \sim 50,000$. Regardless, the precise magnitude of the cavity normalized flux is not of importance for our purposes, only that it saturates. 

     The radially dependent cavity flux is similarly defined:

     \begin{equation}
     \label{eq:radial_bflux}
         \phi_B(r) = \Phi_c(r) / \sqrt{\langle \dot{M}_{\rm tot} \rangle (r/M)^2 c},
     \end{equation}
     where $r$ is the coordinate radius and $M$ is the ADM mass.

    \subsubsection{Four-current and current sheet diagnostic}
In ideal magnetohydrodynamics, the electric field is determined by the magnetic field and the fluid four-velocity, so the inhomogeneous Maxwell equation can be used as a diagnostic to compute the four-current. In Heaviside--Lorentz units, the four-current is given by
    \begin{equation} 
        J^\mu=\nabla_\nu F^{\mu\nu},    
    \end{equation} 
where \(J^\mu\) is the four-current, and \(\nabla_\nu\) is the covariant derivative associated with the spacetime metric \(g_{\mu\nu}\). Antisymmetry of \(F^{\mu\nu}\) gives 
    \begin{equation}
        \label{eq:4curr} 
        J^\mu=\frac{1}{\sqrt{-g}}\left[\partial_t\!\left(\sqrt{-g}F^{\mu t}\right)+\partial_i\!\left(\sqrt{-g}F^{\mu i}\right)\right], 
    \end{equation} 
where \(g\equiv\det(g_{\mu\nu})\) is the determinant of the spacetime metric. The implementation of the current computation was performed and tested in~\cite{Li:2026ngu}.

Here we also compute the magnitude of spatial current measured by observers normal to $t=\rm const.$ slices $j^\mu = \gamma^\mu_\rho J^\rho$ as:

\begin{equation}
    j^2 =\gamma_{ij} j^i j^j = J^\mu J_\mu + (J^\mu n_\mu)^2
\end{equation}
where $i, j = 1,2,3$ are spatial indices, $\gamma_{ij}$ is the spatial metric, and $n_\mu$ is the dual of the normal vector to $t=\rm const.$ slice $n^\mu$.

In addition to the current we also track the electromagnetic invariant $F_{\mu\nu}F^{\mu\nu}=2b^2$, where $b$ is the  magnitude of the magnetic field measured by an observer comoving with the plasma. While in force-free electrodynamics the evolution can make $F_{\mu\nu}F^{\mu\nu}$ negative in current sheets, in ideal MHD $F_{\mu\nu}F^{\mu\nu}>0$~\cite{paschalidis_matching_2013}. However, in current sheets $F_{\mu\nu}F^{\mu\nu}$ becomes very small. In particular, a current sheet is identified by a suppression of the electromagnetic invariant $F_{\mu\nu}F^{\mu\nu}$ and a local maximum of $J^\mu J_\mu$. Thus, we locate current sheets where $M^2 J^\mu J_\mu/F_{\mu\nu}F^{\mu\nu} \gg 1$. The motivation behind this diagnostic is that spatial derivatives of the magnetic field around current sheets become large, while the magnetic field becomes small. In stationary situations the current is (schematically) $J\sim \partial B$, thus a diagnostic of the form $|L\partial B/B|$, where $L$ is some characteristic length scale, gives a measure of whether a current sheet forms. This is what the dimensionless ratio 
\begin{equation}\label{Eq:Dcur_sheet}
    D_{\rm sheet} = \frac{M^2 J^\mu J_\mu}{F_{\mu\nu}F^{\mu\nu}}
\end{equation} 
accomplishes (since for magnetically dominated solutions it schematically is $|M\partial B/B|^2$), where we choose the gravitational mass of the spacetime $M$ as the characteristic lengthscale. After identifying current sheet areas using $D_{\rm sheet}$, we confirm them via images demonstrating magnetic field line reversals.

    \subsubsection{Toroidal and Poloidal Magnetic Pressure}

    We define the toroidal and poloidal magnetic pressure as follows. First, we assume that the toroidal direction is determined by an azimuthal coordinate basis vector $\partial_\varphi$, defined in a standard way through the Cartesian coordinate system centered on the binary center of mass in which our simulations our performed. The toroidal vector in a frame comoving with the plasma (whose four-velocity is $u_{\mu}$) is determined by the projection of $\partial_\varphi$ in a direction normal to $u^\mu$, i.e., $\phi^\mu=\perp^\mu{}_\nu (\partial_\varphi)^\nu$, where $\perp^\mu{}_\nu =\delta^\mu_\nu + u^\mu u_\nu$. The unit toroidal basis vector then is $\hat\phi^\mu = \phi^\mu/\sqrt{\phi^\mu \phi_\mu}.$
    If $b^\mu$ is the magnetic field measured by a comoving observer, then the toroidal component of $b^\mu$ in the comoving frame is just $b^\mu \hat\phi_\mu$. In terms of the center of mass frame quantities, a straightforward calculation shows that the toroidal magnetic field amplitude is given by
    \begin{equation}
         b_{\rm tor}^2 = \frac{(b_\mu (\partial_\varphi)^\mu)^2}{g_{\varphi\varphi} + ((\partial_\varphi)^\mu u_\mu)^2}.
    \end{equation}
    The poloidal component then is $b_{\rm pol}^\mu=b^\mu-(b^\nu\hat \phi_\nu) \hat\phi^\mu$, and its amplitude squared $b_{\rm pol}^2=b^2 -b_{\rm tor}^2$,
    where $b^2 = b_{\mu}b^\mu $. In terms of the Cartesian coordinates adopted in our code, the projections of the magnetic field and velocity 4-vectors, and the $g_{\varphi\varphi}$ components are given by: 

    \begin{equation*}
        b_\mu (\partial_\varphi)^\mu = (-y g_{\mu x} b^\mu + x g_{\mu y}b^\mu),
    \end{equation*}
    \begin{equation*}
        (\partial_\varphi)^\mu u_\mu = (-y g_{\mu x} u^\mu + x g_{\mu y}u^\mu),
    \end{equation*}
    \begin{equation*}
        g_{\varphi \varphi} = y^2 g_{xx} + x^2 g_{yy} - 2xy g_{xy}.
    \end{equation*}

    To obtain the magnetic pressure in Gaussian units, one divides $b_{\rm tor}^2$ and $b_{\rm pol}^2$ by $8\pi$.
    The above decomposition ensures that the toroidal and poloidal components of the magnetic pressure are those measured by an observer comoving with the plasma, because both $b^\nu_{\rm tor} u_\nu=0$, and $b^\nu_{\rm pol} u_\nu=0$. 
    
\section{Results} \label{sec:results}

In Figure~\ref{fig:3d_plot}, we depict the BBH-CBD system in our simulation via a volume rendering of the rest-mass density at $t/M = 46339.2$, after approximately $\sim 47$ orbits. In the top left panel, we plot the entire CBD on a color scale. In the bottom right panel, we zoom into the cavity region, indicate the cavity region, the minidisks, and describe the horizon magnetic fields with white streamlines. The cavity is elliptical and asymmetric. A high density accretion stream is approaching the `left' BH from the inner edge of the CBD. This is also the BH closer to the CBD. That tidal stream is feeding the minidisk around the BH. 

In the rest of this section, we present key analysis and diagnostics of the simulation, including the mass accretion rates, horizon magnetic flux, cavity magnetic flux, Hill sphere (minidisk) masses, cavity currents, relative toroidal and poloidal magnetic pressures, and fluid temperature.

\subsection{Multiphase Relaxation of Accretion and Cavity Flux} \label{subsec:multiphase}

\begin{figure*}
    \centering
    \includegraphics[width=1\linewidth]{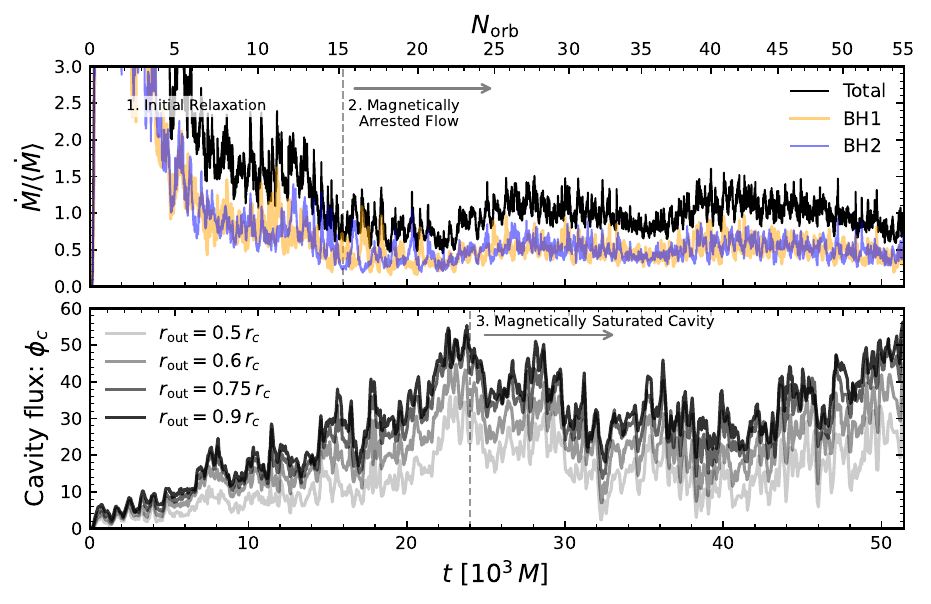}
    \caption{Top: rest-mass accretion rate onto each of the horizons (blue and gold) and their sum (black) normalized by the average from $25 < t/10^3 M < 40$. Bottom: total vertical magnetic flux threading the $z=0$ plane integrated from the BBH center of mass out to multiple $r_{\rm out}$ radii within the cavity region (as defined by Eq.~\ref{eq:cavity_flux}). The accretion rate goes through three phases: 1. $t < 16,000\, M$; the fluid initially relaxes onto the binary, 2. $t > 16,000 \, M$; the BBHs begin to host magnetically arrested minidisks, 3. $t > 24,000 \, M$; the cavity flux is saturated (see bottom panel) and the accretion rate settles onto a new average with a long-term modulation. Throughout phases 1 \& 2 the cavity magnetic flux is increasing. It then saturates at $\sim 24,000 \, M$, before oscillating around $\phi_c \sim 35$ throughout phase 3; this is what we call the magnetically saturated, relaxed accretion flow. The BBH-CBD has entered a dual horizon magnetically arrested disk state (DH-MAD).}
    \label{fig:mdot_cavity_flux}
\end{figure*}

\begin{figure}
    \centering
    \includegraphics[width=1.05\linewidth]{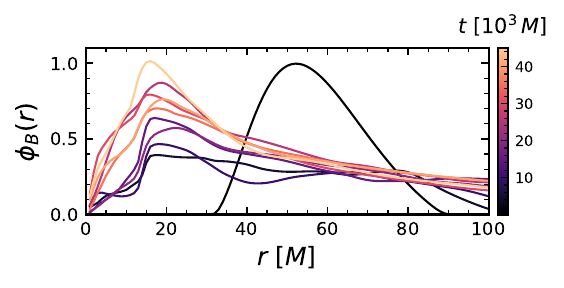}
    \caption{Radial profile of the cumulative magnetic flux threading the $xy$ plane from $t/M = 0 \text{ to } 45,000$ plotted on a colormap where brighter colors are later in the simulation. The flux within the cavity peaks at $\phi_B \sim 1$ at $r/M \sim 15$ for the latest time plotted.}
    \label{fig:radial_bflux}
\end{figure}

In Figure~\ref{fig:mdot_cavity_flux}, we plot the rest-mass accretion rate normalized to the average, $\langle \dot{M} \rangle$, for $25-40\times 10^3 M$ (top panel) and the approximate cavity magnetic flux ($\phi_c$, bottom panel)(see \ref{subsec:b_flux} for a description of the calculation). In the top panel, we plot the individual BH accretion rates with blue and gold lines and the total accretion rate with a black line. The total rest-mass accretion rate has three general phases: 1. initial relaxation ($t/M < 16,000$), 2. dual-horizon magnetically arrested accretion flow ($t/M > 16,000$), and 3. magnetically saturated cavity with continued magnetically arrested horizon accretion flow ($t/M > 24,000$). The beginning of the second phase is initially characterized by persistent minidisks and magnetic reconnection events that form flux bundles which are expelled from the horizon vicinity -- previously discussed in~\cite{manikantan_magnetically_2025}.

The subsequent evolution is characterized by the cavity magnetic flux saturating and continued flux eruptions from vicinity of the BH horizons. In the bottom panel, we plot the cavity flux, $\phi_c$, as measured out to different radii from the center of mass. $r_c$ is the cavity coordinate radius (measured from the center of mass) defined as $r_c = d$, where $d$ is the time-dependent BBH separation. Hence, an initial separation of $d/M = 30$ implies a cavity radius of $r_c/M = 30$. This is not exact, as the cavity wall is not well defined nor is it circular. Therefore, we define multiple radii for the $r_{\rm out} = {0.5, 0.6, 0.75, 0.9} \, r_c$ plotted in increasingly darker gray lines. The last two $r_{\rm out}$ are nearly the same, suggesting that most of the vertical magnetic flux is contained within $0.75 \, r_c$.

During the first two phases, the cavity flux is increasing to a global maximum $\phi_c \sim 55$ at $t/M \simeq 24,000$. This appears to be driven in part by flux eruptions from the BH horizons -- drops in the horizon flux are matched directly by increases in the cavity flux. However, whether this flux persists within the cavity and for how long is not clear, as we are not using particle tracers. Additionally, how early and rapidly this process fills the cavity with magnetic flux is expected to be sensitive to the seeded magnetic field. This still needs to be tested in full GRMHD. However, flux build up in the cavity that eventually reaches a maximum is an expected outcome of evolutions that reach steady state. Whether this maximum corresponds to a saturated or magnetically arrested state, depends on the amount of magnetic flux and its orientation with respect to the cavity wall (as we discuss later). 

During the magnetically saturated cavity phase, the rest-mass accretion rate settles at an average of $\dot{M} \sim 1$ (by design, as we have normalized the signal to the average in this phase) and the cavity flux fluctuates around $\phi_c \sim 35$. The total rest-mass accretion rate undergoes a long-term modulation with a period of about $T\sim14,000 \, M$ (see peaks at about $26,000$ and $40,000 \, M$). However, with only two cycles this quasiperiodicity is only a preliminary result. We do notice that this period corresponds to the orbital angular velocity of the plasma near the outer disk edge of our accretion disk. However, longer simulations are required to make any strong connections. In Appendix~\ref{sec:app_lump}, we describe further analysis of the CBD and the harmonic modes in its density distribution.

In Figure~\ref{fig:radial_bflux}, we plot the radial dependence of the cumulative cavity vertical flux (see Eq.~\ref{eq:radial_bflux}). The initial magnetic flux through the $z=0$ surface is in black and peaks at $r/M \sim 50$, and increasing times are brighter in color (see the color bar to the right). At its maximum, which is also at the latest time plotted, the cavity flux peaks at $\phi_{B} \sim 1$ at $r/M \sim 15$. Typically, values of $\phi_{B} >> 1$ in single accreting BHs  indicate a magnetically arrested state at the BH horizon~\cite{tchekhovskoy_mad_2011}. The precise value of $\phi_{B}$ which indicates a magnetically arrested cavity in the binary case needs further study with initial conditions of varying net vertical flux in the initial disk. We normalize the flux in Figure~\ref{fig:radial_bflux} by $c$ (see Equation~\ref{eq:radial_bflux}) to compare with previous works~\citep[e.g. Ref.][who measure $\phi_B > 10$ in their cavity]{most_magnetically_2024}. However, a direct treatment of magnetic and ram pressure balance suggests that the radial velocity component $v_r$ is a more appropriate quantity in the normalization factor to identify a ``magnetically arrested'' flow. Our choice of using $c$ in the normalization is therefore used to provide a lower bound on the value of the normalized flux. Furthermore, even if there is an azimuthally average pressure balance between ram and magnetic pressure, this does not automatically imply choked accretion. This is because binary accretion (for near equal mass BHs) is inherently non-axisymmetric, which can allow the accretion flow to `bypass' the magnetic field barrier and for accretion to continue, even in a so-called magnetically arrested state. Given these qualifications, we use ``magnetic flux saturation" in the remainder of the paper to mean that the magnetic flux has peaked and does not increase any further. 

\subsection{Accretion Rate and Dimensionless Horizon Flux Periodicity} \label{subsec:mdot}

\begin{figure*}
    \centering
    \includegraphics[width=1\linewidth]{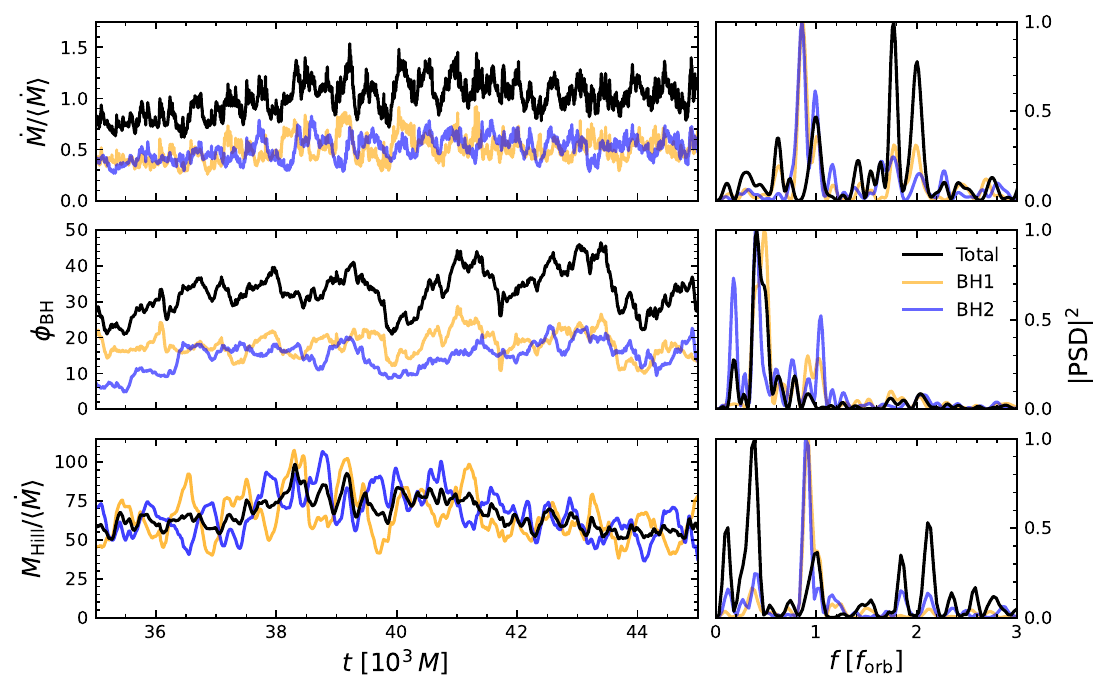}
    \caption{Top left: rest-mass accretion rate onto each BH (gold and blue) and their sum (black) all normalized by the average of the summed accretion rate for $ 35 < t/10^3 M < 45$. Middle left: dimensionless horizon flux parameter for each BH and their sum. Bottom left: time evolution of the mass contained within the Hill sphere of each BH and their sum normalized by the average individual and summed mass accretion rates, respectively. This quantity has units of time and describes an inflow time for the minidisks in units of $M$. Right column: Fourier transform of the signals on the left where long term modulations (periods of $T > 5,000\,M$) have been subtracted out. The rest-mass accretion rate demonstrates periodicity at $f \sim 0.9 f_{\rm orb}$ for each BH and $f \sim 1.8f_{\rm orb}$ for the sum of the two. This suggests that accretion alternates onto the BHs. The total dimensionless horizon flux is periodic at $f\sim 0.4 f_{\rm orb}$ across all three timeseries. The Hill sphere mass of each BH is periodic at $f \sim 0.9 f_{\rm orb}$ and out of phase with the other BH, consistent with the mass accretion rates. The total signal is periodic at $f \sim 0.4 f_{\rm orb}$.}
    \label{fig:all_ts}
\end{figure*}

In Figure~\ref{fig:all_ts}, we analyze the periodicity of the rest-mass accretion rates (top), the horizon dimensionless fluxes (middle), and the mass contained within the Hill spheres of each BH (bottom). In the left column, we plot the timeseries of each quantity and in the right column we plot the power spectral density (PSD) of the Fourier transform. Before performing the Fourier analysis, we subtract the running average from the signal to remove the direct current (DC) component and reduce spectral leakage from slow trends. 

In the top left panel we plot the rest-mass accretion rate onto each BH (orange and blue lines) and their sum (black lines) normalized by the average of the sum for the shown time period. The PSD shows a dominant accretion periodicity of $f \sim 0.9 f_{\rm orb}$ for the individual timeseries with secondary peaks at $f \sim 1.8 f_{\rm orb}$.  The individual BHs accrete out of phase, so when summed the $f \sim 0.9f_{\rm orb}$ harmonics interfere destructively, leaving only the modulation at $f \sim 1.8 f_{\rm orb}$ for the total signal.

In the middle left panel of Figure~\ref{fig:all_ts}, we plot the dimensionless horizon flux parameter ($\phi_{\rm BH}$; equation~\ref{eq:phibh}) for each horizon (orange and blue) and the sum of the two signals (black). We can identify flux eruption events (originating from the vicinity of the horizons) as rapid drops in the individual signals (see, e.g., $t/10^3 M \simeq 40, 42, 43.5$). The PSD of the signal reveals a clear dominant periodicity at $f \sim 0.4 \, f_{\rm orb}$.

Finally, in the bottom row we describe the rest-mass contained within the Hill sphere of each BH. The Hill spheres have clear alternating pattern -- when one BHs Hill mass peaks, the other is at a minimum. The individual Hill masses have a periodicity at $f \sim 0.9 f_{\rm orb}$, which is consistent with the accretion rate periodicity onto the individual BHs, and hence the frequency of matter stripping from the disk inner edge. The total Hill sphere mass is periodic at $f \sim 0.4 f_{\rm orb}$. The alternating individual timeseries destructively interfere like the individual mass accretion rate timeseries. However, in this case, the remaining periodicity is at $f \sim 0.4 f_{\rm orb}$, which is the secondary frequency in the individual timeseries.

The measured periodicity in our total rest-mass accretion rate is slightly different to previous studies of MHD accretion onto nonspinning, equal-mass, quasicircular BBHs, which usually report a periodicity in the range $f \sim 1.4-1.6 f_{\rm orb}$~\cite{bright_minidisk_2023, paschalidis_minidisk_2021, cattorini_grmhd_2024, manikantan_effects_2025}. The key change in this study is the large initial separation ($d=30 \, M$) and long integration time ($t > 20,000 \, M$) which allows for $> 20$ orbits at a near constant separation, for the accretion flow to relax, and for magnetic flux to saturate the cavity. This has not previously been done in full GRMHD. The periodicity of the rest-mass accretion rate is expected to manifest in the thermal UV/X-ray spectrum of SMBBHs, as the hot accretion flow near the horizons is modulated at this periodicity~\cite{bogdanovic_electromagnetic_2022}. 

Furthermore, the $f \sim 0.4 f_{\rm orb}$ variability of the total dimensionless horizon flux establishes a periodicity in a new potential EM counterpart for SMBBHs. The driver of this periodicity is still unclear. We find that rapid drops in $\phi_{\rm BH}$ are primarily driven by magnetic flux eruptions from the horizon. These are sites of magnetic reconnection and can drive nonthermal flaring~\cite{dexter_sgr_2020, scepi_sgr_2022}. However, we have not been able to identify discrete flux eruption events with the periodicity of the timeseries throughout the entire evolution. This may be due to the fact that eruptions are 3D effects, and not all them may lie in the equatorial plane we mostly analyze.

If the $\phi_{\rm BH}$ periodicity was instead was driven solely by accretion, we would expect the horizon flux to have the same periodicity as the individual rest-mass accretion rate and the Hill sphere masses ($f \sim 0.9 f_{\rm orb}$). However, the individual $\phi_{\rm BH}$ are variable at $f \sim 0.4 f_{\rm orb}$, which is half as frequent as the accretion onto the horizons, suggesting that it takes up to $2$ accretion episodes to saturate the horizons with magnetic flux. It is unclear what causes this periodicity in horizon magnetic flux. Simulating different BBH parameters (e.g. spins, mass ratio) would be necessary to test the robustness of these periodicities. Although we expect that the new DH-MAD accretion mode we report should be robust against binary parameters, we plan to test this in forthcoming work.

\subsection{Cavity and Minidisk Evolution} 
\label{subsec:cavity}

\begin{figure*}
    \centering
     \includegraphics[width=1\linewidth]{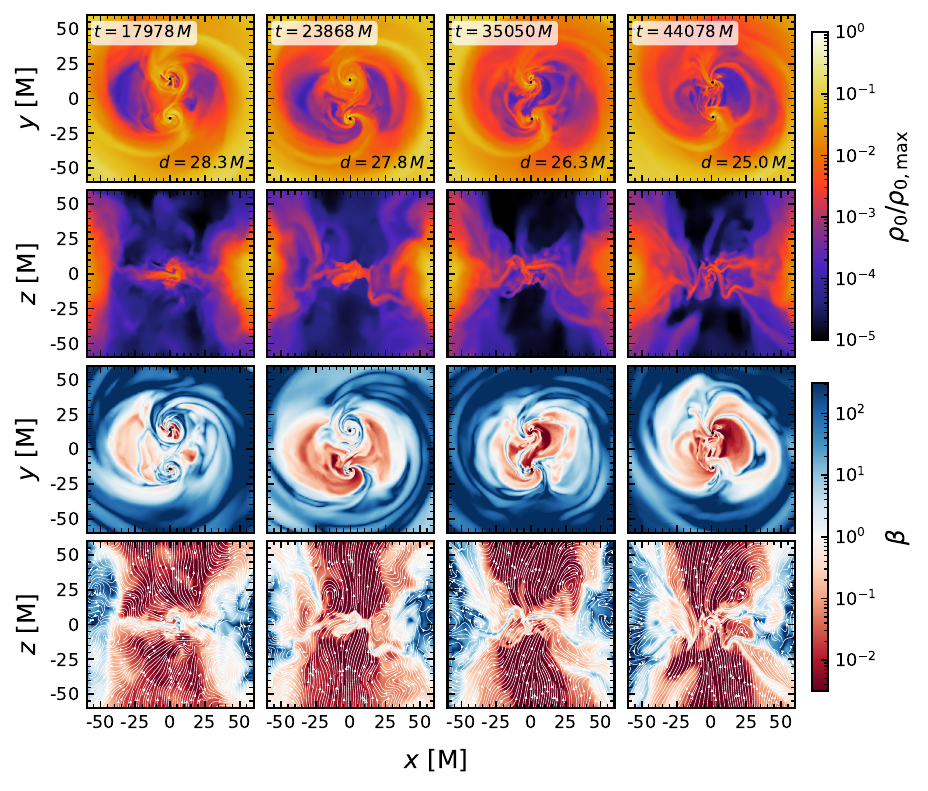}
    \caption{$xy$ and $xz$ slices with magnetic fields of our BBH-CBD system at four times: 1. $t/M \sim 17,978$, when hosting magnetically arrested minidisks, 2. $t/M \sim 23,868$, at peak cavity magnetic flux, 3. $t/M \sim 35,050$, at a long-term minimum in relaxed phase accretion rate (see Figure~\ref{fig:mdot_cavity_flux}), and 4. $t/M \sim 44,078$, late in the evolution. In the first two rows we plot the rest-mass density of the gas normalized by the initial maximum in the CBD ($\rho_0/\rho_{0, \rm \, max}$) and in the last two rows the plasma beta parameter ($\beta\equiv P_{\rm gas}/P_{\rm mag}$). In the first and third we plot an $xy$ slices of the system (face-down) and in the second and fourth rows the $xz$ slices (side-on) with the magnetic field lines overplotted in white. The cavity is elliptical, lopsided, and dominated by magnetic pressure at later times.}
    \label{fig:rho_beta}
\end{figure*}

\begin{figure*}
    \centering
    \includegraphics[width=0.8\linewidth]{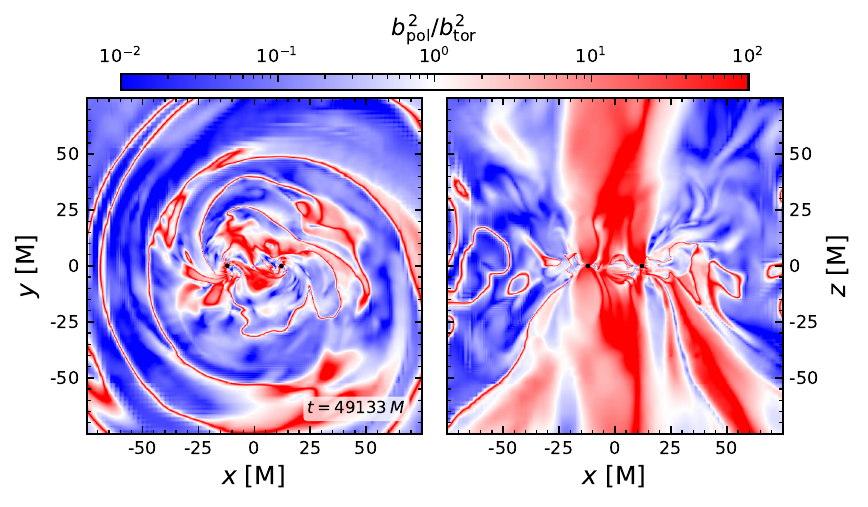}
    \caption{Ratio of the poloidal ($b^2_{\rm pol}/8\pi$) and toroidal ($b^2_{\rm tor}/8\pi$). Red (blue) indicates poloidal (toroidal) pressure dominates ($b^2_{\rm pol}/b^2_{\rm tor} > 1$ and $b^2_{\rm pol}/b^2_{\rm tor} < 1$, respectively). Near the BHs and in the jet regions poloidal pressure dominates but in the bulk of the cavity and the CBD the toroidal pressure dominates. Toroidal magnetic fields increase magnetic tension at the cavity wall and can suppress the growth of instabilities.}
    \label{fig:bpol_over_btor}
\end{figure*}

In Fig.~\ref{fig:rho_beta} we describe four characteristic times in the DH-MAD phase. In the top two rows we plot a color map of the rest-mass density normalized by the initial maximum in the accretion disk ($\rho_0/\rho_{0, \rm max}$). The normalization corresponds to
\begin{equation}
    \rho_{0, \rm \, max} = 4.42 \times 10^{-11} \biggl(\frac{\langle\dot{M}\rangle}{0.1 \rm \, \dot{M}_{\rm edd}}\biggl) \biggl(\frac{10^7 M_{\odot}}{M_{\rm tot}}\biggl)^{-1} \rm \, g/cm^3.
\end{equation}
In the bottom two rows we show a color map of the plasma beta parameter ($\beta \equiv P_{\rm gas}/P_{\rm mag}$) on a diverging color scale where red (blue) indicates magnetic (gas) pressure dominated. We plot each quantity in the $xy$ (``face-on") and $xz$ (``edge-on") planes of the system. In the $xy$ slices, we also indicate the BH horizons with black disks whose radius corresponds to the actual coordinate radius of the horizon -- in all snapshots we have chosen times where the BHs are on the y-axis ($x=0$). We also plot the magnetic field lines with directed white streamlines in the plasma beta $xz$ slices. 

In the first column, we choose the time $t/M = 17,978$ while the minidisks are magnetically arrested (MAM phase). The inner edge of the circumbinary disk is at $r/M\sim 40$; although at this point in the evolution, the cavity is somewhat elliptical (with the major axis perpendicular to the binary axis in this snapshot). There are two high-density streams ($\rho_0/\rho_{0, \rm max} > 10^{-1}$) falling towards each BH, which are at $y/M \simeq \pm 14$. The BHs host equally high density minidisks (again, $\rho_0/\rho_{0, \rm max} > 10^{-1}$). Perpendicular to the BHs are low density areas ($\rho_0/\rho_{0, \rm max} < 10^{-3}$) that approximately trace the cavity boundary. The `top' BH's ($y/M = 14$) minidisk is undergoing an active magnetic flux eruption event, as indicated by the areas of low density fluid in the immediate vicinity of the BH.\footnote{Poloidal magnetic fields on the BH horizon squeeze at the equator to produce current-sheet-like structures. The field then reconnects through this current sheet to form a magnetic flux bundle that buoyantly rises away from the horizon; this is a characteristic of a magnetically arrested disk~\cite{tchekhovskoy_mad_2011, manikantan_magnetically_2025, most_magnetically_2024, mckinney_mad_2012})}

In the third row of the first column, the $xy$ plasma $\beta$ slice, the magnetic flux eruption is more evidently described by regions of high plasma $\beta$ ($\equiv$ high fluid magnetization). The minidisks around each BH are predominantly gas pressure dominated ($\beta > 1$). However, the low density cavity on the negative $x$-axis (left of the BHs) is magnetic pressure dominated ($\beta < 1$) . The right side of the cavity is gas pressure dominated ($\beta > 1)$, although there is a magnetic flux bundle `floating' in the cavity as indicated by the low gas density and $\beta < 1$ at around $y/M = -5$ and $x/M = 15$.

In the second and fourth rows, we show vertical ($xz$) slices of the system at a time that the slices cut through the system perpendicular to the BHs. The gas density (second row, first column) is low ($\rho_0/\rho_{0, \rm max} < 10^{-3}$) in the majority of the slice shown. At the left and right boundaries ($x = \pm 50$) we see the inner edge of the CBD ($\rho_0/\rho_{0, \rm max} \sim 10^{-2}$). The dual-jet launched by the BBHs (not pictured in this slice) evacuate the polar regions of dense gas, leaving behind low-density, highly magnetized gas. We describe the polar region's high magnetization in the fourth row where $\beta \ll 1$ below and above the BHs. On top of the color map, we plot the magnetic field lines with white streamlines. The inner edge of the CBD is gas pressure dominated ($\beta>1$) and approaches equipartition at the boundary between disk and cavity. The majority of the cavity ($r/M < 40$ and $\rho_0/\rho_{0, \rm max} < 10^{-2}$) is magnetic pressure dominated ($\beta < 1$) with filaments of gas pressure dominated fluid making its way towards the center.

In the second column we plot the BBH system when the cavity flux $\phi_c$ is at its first peak at $t/M = 23,868$. This peak is most obviously reflected in the $xy$ slice of the fluid plasma beta (third row) where the cavity is now magnetic pressure dominated ($\beta > 1$) throughout, except for the gas that is falling from the inner edge of the CBD. Additionally, notice that only one of the BHs hosts a large minidisk. At this point in the evolution, an $m=1$ azimuthal density mode has developed in the CBD and accretion alternates onto the BHs~(see App.~\ref{sec:app_lump} for an extended discussion on the evolution of the CBD). The fourth row of this snapshot also reveals similar filament structure in the cavity, an even lower fluid plasma beta (indicating even greater magnetic pressure domination), and more ordered, vertical magnetic fields threading the $z=0$ plane. Additionally, there are magnetic field loops at the interface between the CBD and cavity/jet region, for example, at $x,y=(-25,10)$, which possibly developed at the shear layer of the fast-moving jet and slow-moving wind~\cite{sironi_reconnection-driven_2021}. 

In the third column, we plot the BBH at a minimum in its mass accretion rate at $t/M = 35,050$ (see Fig.~\ref{fig:mdot_cavity_flux}). Neither BH hosts a persistent minidisk and accrete gas pressure dominated streams directly from the CBD. Finally, the vertical magnetic field is still highly ordered within the cavity and filaments of gas pressure dominated fluid enter the cavity. 

In the final column of Fig.~\ref{fig:rho_beta}, we plot the system at $t/M\sim 44,000$, which is about $45$ orbits in the evolution. The system has settled into a quasi-steady state and retains many of the features of the previous time: the asymmetric, elliptical cavity, and intermittent minidisks. Near the BHs, the fluid is strongly dominated by magnetic pressure ($\beta \sim 10^{-2}$), and the cavity remains dominated by magnetic pressure to $r/M \simeq 50$ at certain azimuths. There is also a temporary gas-pressure dominated filament that separates the $\beta \sim 10^{-2}$ and $\beta \sim 10^{-0.5}$ regions. Notably, even though the vertical magnetic flux in the cavity is saturated, we do not observe any flux eruptions from the cavity into the CBD. 

To further understand the structure of the magnetic field within the cavity, in Fig.~\ref{fig:bpol_over_btor} we plot the ratio $b^2_{\rm pol}/b^2_{\rm tor}$ to describe the relative strength of the toroidal to poloidal magnetic field pressure. We see that near the BHs and in the funnel below and above the BBH orbital plane the poloidal magnetic pressure dominates (red), as is expected from the collimated incipient jets launched by the BHs. However, right outside the orbit of the BHs ($r/M \gtrsim 20$) the fluid is toroidal magnetic pressure dominated (blue). It is possible that the strong toroidal field in the cavity explains why we see no flux eruptions from the cavity into the CBD, which were previously reported in~\cite{most_magnetically_2024, wang_bmad_2025}. In particular, the strong toroidal field in our simulations reduces the growth of interchange perturbations that vary predominantly in the azimuthal direction ($\hat{\varphi}$). In the local planar limit, the stabilizing magnetic-tension term scales as $(\boldsymbol{k}\cdot\boldsymbol{B})^2$, where $\boldsymbol{k}$ is the wavevector of any perturbations in the gas density and $\boldsymbol{B}$ is the magnetic field vector \citep[Chapter~X, \S97]{Chandrasekhar1961}. A higher magnetic-tension term makes it more difficult to develop instabilities, such as Rayleigh-Taylor or magnetic interchange. Therefore, a radial displacement that varies in azimuth and is nearly uniform with height would bend a coherent toroidal field (hence $\boldsymbol{k} \, || \, \boldsymbol{B} \, || \hat \varphi$) while producing little bending of a predominantly vertical field. This mechanism could reduce the growth of such instabilities (or `fingers') at the cavity wall, and prevent flux bundles from erupting into the CBD. Although, oblique modes can still grow if their wavevectors are nearly perpendicular to the local field (i.e. $\boldsymbol{k} \cdot \boldsymbol{B} = 0$).  

The treatment of accretion at the horizon may also affect how vertical flux accumulates and is distributed in the cavity. The Newtonian sinks of Ref.~\cite{most_magnetically_2024} cannot self-consistently treat horizon-magnetic field interactions, whereas our MHD simulations in full GR resolve accretion through the individual horizons. Additionally, the simulations presented in Refs.~\cite{most_magnetically_2024, wang_bmad_2025} have an initial magnetic field that is $\sim 2.5$ times stronger than in the simulation we present here (their $\beta_{\rm ini}=20$ as compared to our $\beta_{\rm ini}=52$). Although, the exact initial plasma beta parameter may be less important than the total magnetic flux seeded in the CBD. Further exploration of BBH and CBD parameters is necessary to determine when DH-MAD or BMAD accretion states may arise.

\subsection{Current sheets}
\begin{figure}
    \centering
    \includegraphics[width=1\linewidth]{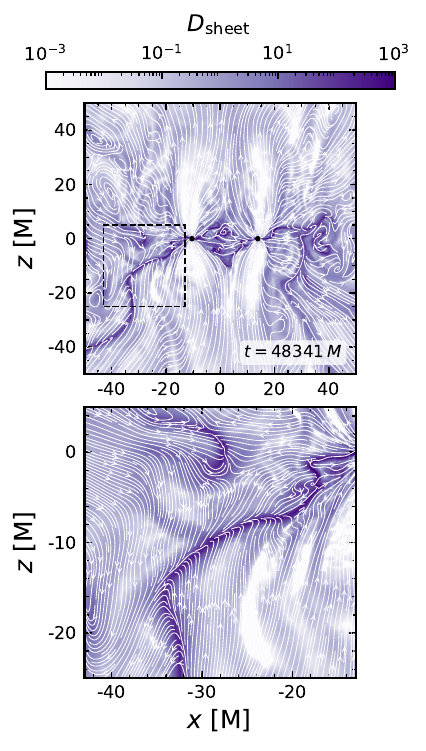}
    \caption{Current sheet diagnostic ratio $D_{\rm sheet}\equiv M^2 J^2/ 2b^2$ (equation~\eqref{Eq:Dcur_sheet}) and field lines corresponding to the magnetic field measured by a normal observer.}
    \label{fig:current}
\end{figure}

In Figure~\ref{fig:current}, we plot the current sheet diagnostic ratio $D_{\rm sheet}$ of Eq.~\eqref{Eq:Dcur_sheet} on a color scale where lighter colors mean small value of $D_{\rm sheet}$, and darker colors indicate large values of $D_{\rm sheet}$. We also overlay the magnetic field lines corresponding the magnetic field by a normal observer. We see that the bulk of the system is lighter colors ($D_{\rm sheet}<1$), especially in the jet funnel of the BBH. However, close to the horizons, we see $ D_{\rm sheet} > 1$, suggesting that current sheets develop near the BHs. This is unsurprising~\cite{ripperda_black_2022}. The combination of ordered magnetic fields that thread the horizon and reconnection that produces magnetic flux bundles both point to the formation of current sheets at the horizon-scale. 

Interestingly, there is also a dark region of $D_{\rm sheet}\gg 1$ extending from the `left' BH, through the cavity, and into the wind region directly below the bulk of the CBD. In the bottom panel, we plot a zoomed-in view of this current-sheet-like structure. In addition to $D_{\rm sheet} > 10^2$, the magnetic field lines reverse direction, which indicate that this is a current sheet where magnetic reconnection took place. Such reconnection sites can accelerate particles to non-thermal distributions and power high-energy phenomena. We do caution that the resolution in this simulation at $20 < r/M < 40$, is $dx \simeq 0.5M$, and the thickness of these current sheets are $dx \simeq 2M$, so they are not really resolved. 

\subsection{Flow Temperature}
\begin{figure}
    \centering
    \includegraphics[width=0.75\linewidth]{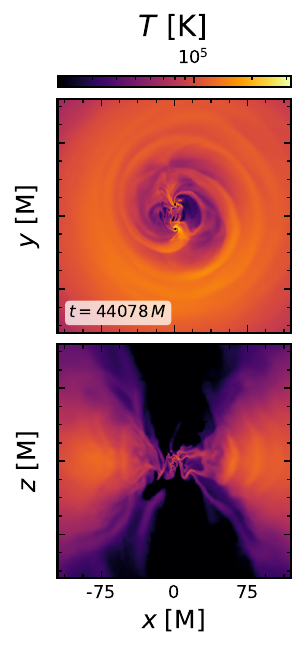}
    \caption{Fluid temperature in degrees Kelvin assuming a radiation pressure dominated accretion flow (consistent with our choice of $\Gamma = 4/3$), $M_{\rm tot} = 10^7 \, M_{\odot}$, and $\dot{M}_{\rm tot} = 0.1 \dot{M}_{\rm edd}$ with a $10\%$ efficiency of the accretion flow. Top panel ($xy$ slice): the bulk of the CBD has fluid temperature $10^5 \, K$ at the midplane; this is not the temperature at the photosphere. The fluid temperature also captures the spiral density waves. The cavity has a lower average temperature ($T_{\rm cav}< 10^5$) but the high density streams have a higher temperature than the CBD. Bottom panel ($xz$ slice): the temperature is highest at the midplane and falls off as you reach the vertical extent of the accretion disk. The jet regions (below and above the cavity) have very low fluid temperature.     }
    \label{fig:cavity_temp}
\end{figure}

\begin{figure}
    \centering
    \includegraphics[width=\linewidth]{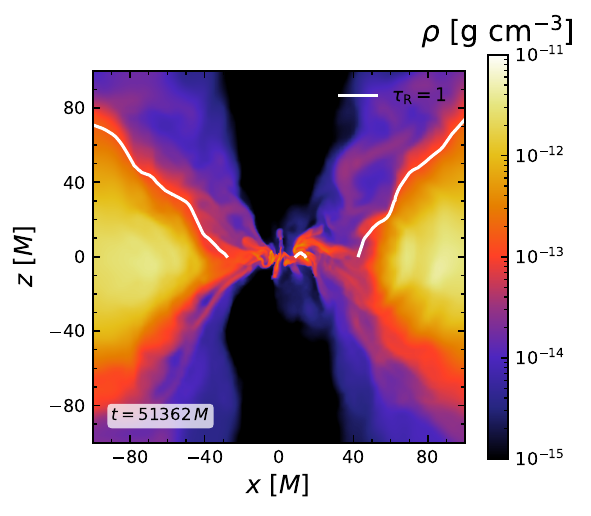}
    \caption{We plot an $xz$ density slice of the simulation assuming a $10^7 M_{\odot}$ SMBBH accreting at $10\%$ of the Eddington accretion rate. On top, we plot the surface of optical depth of unity ($\tau=1$) as viewed by an observer directly above the SMBBH ($\theta=0$), assuming Rosseland mean opacity and terminating the integration at $z=0$ if $\tau=1$ is not found. For $|r/M| < 40$, the cavity region, the fluid is optically thin.}
    \label{fig:tau1}
\end{figure}

We calculate the fluid temperature from the gas pressure in our simulation assuming a mix of radiation and gas pressure,
\begin{equation}
    P = \frac{1}{3}aT^4  + \frac{\rho_0}{\mu m_u}kT,
\end{equation}
where $P$ is the gas pressure measured in our simulation $a=7.5675 \times 10^{-15} \rm \, erg \, cm^{-3} \, K^{-4}$ is the radiation constant, $\rho_0$ is the rest-mass density, $\mu$ is the mean molecular weight (taken equal to $1/2$ assuming pure Hydrogen), $m_u$ the atomic mass unit, $k$ is the Boltzmann constant, and $T$ is the fluid temperature. Given our choice of $\Gamma = 4/3$ in the equation of state, and that the radiation pressure term dominates, the temperature can be simplified to $T = (3P/a)^{1/4}$. Although, we do not make this assumption when computing the temperature.

To give physical units to the gas density and pressure, we apply the following procedure:
\begin{enumerate}
    \item Choose a binary gravitational mass and target average mass accretion rate. In this case, we choose $M= 10^7 \, M_{\odot}$ and an accretion rate of $\langle\dot{M}\rangle = 0.1 \dot{M}_{\rm edd}$. We also assume a radiative efficiency $\eta=10\%$. For our choice of $\Gamma=4/3$, this is an appropriate radiation-dominated accretion flow.  
    \item Restore units to the simulation's measured mass accretion rate by multiplying by $c^3/G$. Also, restore units to the rest-mass gas density ($c^2 / Gr_g^2$) and the gas pressure ($c^4 / Gr_g^2$) where $r_g \equiv GM/c^2$.
    \item Find the ratio between the code mass accretion rate and the target mass accretion rate. 
    \item Multiply the density and the gas pressure by the above ratio to obtain quantities with physical units and scale corresponding to the chosen binary mass and mass accretion rate.
\end{enumerate}

The Eddington mass accretion rate is given by:
\begin{equation}
    \dot{M}_{\rm edd} = L_{\rm edd}/\eta c^2 = \frac{4 \pi G M m_p c}{\sigma_T \eta c^2},
\end{equation}
where $m_p$ is the proton mass and $\sigma_T$ the Thompson scattering cross section.

In Fig.~\ref{fig:cavity_temp}, we plot the fluid temperature on $xy$ and $xz$ slices. The peak temperature at the midplane is around $T\sim 2\times 10^5 K$ occuring in the high density streams that fall asynchronously onto each BH. The cavity itself ($r/M < 30$) is thermally `cold', although it can still produce high-energy radiation due to magnetic flux eruptions. The bulk of the CBD has temperature $T \sim 10^5 K$ (consistent with previous calculations in full GR at much smaller separations, see e.g.~\cite{gold_accretion_2014,Gold:2013zma}) which corresponds to a black body emission peak wavelength of about $30\, \rm nm$. This would be extreme UV in the source frame, and in the observer frame will be redshifted by a factor of $(1+z)$, where $z$ is the cosmological redshift. In the $xz$ slice of Fig.~\ref{fig:cavity_temp}, there is decreasing temperature as you move away from the midplane of the CBD. The observed blackbody temperature would depend on the temperature of the photosphere -- optical depth of unity ($\tau=1$) -- which is dependent on both the viewing angle and the absorption/emission mechanisms you include. 

In Figure~\ref{fig:tau1}, we plot the surface of optical depth unity (i.e., $\tau=1$) assuming a $10^7 M_\odot$ SMBBH accreting at $10\%$ of the Eddington accretion rate, viewed from directly above the system (at the plane $z/M=100$). We assume fully ionized, pure hydrogen gas and use the Rosseland mean opacity, which is a harmonic average of free-free absorption and electron scattering opacities~\cite{rybicki_lightman}. In this case, electron scattering dominates the opacity and is effectively equal to the Rosseland mean. The purpose of this calculation is to demonstrate that the cavity is optically thin as compared to the bulk of the CBD. Predicting the precise emission from this system requires careful general-relativistic ray-tracing (GRRT), which will be the subject of upcoming work.

\subsection{Jet Luminosity}

Single, nonspinning BHs do not typically launch jets or Blandford-Znajek-type outflows~\cite{blandford_znajek}. When in a binary, however, the angular momentum needed to twist up the magnetic fields and launch such a jet can be provided by the orbital motion of the BBH~\cite{palenzuela_dual_2010}. That is the case in this simulation. We measure an average jet luminosity, measured by the Poynting luminosity through a surface at $r/M = 200$, of $L_{\rm Poyn}/\dot M=0.01$ (in other words, jet efficiency $\eta = 1\%$). This is a factor of $\sim10$ below the value in previous studies in full GR with non-spinning black holes at separations of $10M$ (see e.g.~\cite{farris_binary_2012}) but comparable to those at $20M$ (see e.g.~\cite{bright_minidisk_2023, manikantan_effects_2025}. Assuming that the power comes from a Blandford-Znajek type effect, the luminosity scales like $B^2 v^2$~\cite{palenzuela_dual_2010}, with $v^2\sim (r/M)^{-1}$ the binary orbital velocity, we do expect an increase of the Poynting luminosity with decreasing orbital separation. Therefore, a factor of 2-3 compared to Ref.~\cite{farris_binary_2012} can be accounted for by the different orbital separation. The remaining factor most likely is due to the different initial vertical magnetic flux in the adopted disks.

\section{Summary and Discussion} \label{sec:summary}

In this work, we have described the longest MHD simulation of magnetized accretion onto BBH in full GR to-date. We initialize a quasicircular, nonspinning, BBH at $d/M \sim 30$ which then inspirals to $d/M \sim 24$ during the simulation which is evolved to $t/M \sim 52,000$ with a highest resolution of $\Delta x = M/64$ ($\sim 64$ cells across each BH horizon). 

The BBH-CBD system goes through three phases: 1. initial relaxation, 2. persistent magnetically arrested minidisks~\cite{manikantan_magnetically_2025}, 3. relaxed magnetically saturated cavity accretion. In the final two phases, the BBH-CBD system has entered a dual horizon-MAD state. In this work, we describe this final state and we summarize our findings below:

\begin{enumerate}
    \item The magnetic flux in the cavity saturates after $\sim 23$ orbits and coincides with a new relaxed phase of accretion, which we call dual horizon MAD, because magnetic flares originate only from the vicinity of the BH horizons.
    
    \item The total (individual) accretion rate onto the BHs is modulated at a periodicity of $f\sim 1.8 f_{\rm orb}$ ($f\sim 0.9 f_{\rm orb}$).
    
    \item The dimensionless horizon magnetic flux is modulated with a robust periodicity of $f \sim 0.4 f_{\rm orb}$, suggesting that quasiperiodic nonthermal flaring could be a smoking gun EM counterpart to inspiraling SMBBHs with a magnetically saturated cavity. 
    
    \item We do not observe magnetic flares from the cavity as in~\cite{most_magnetically_2024}. Toroidal magnetic pressure dominates in the cavity, likely suppressing the growth of instabilities at the cavity wall and any flux eruptions into the CBD.
    
    \item Minidisks do not persist around each BH, accretion streams briefly circularize before plunging, and magnetic flux eruptions dominate the near horizon dynamics.
    
\end{enumerate}

Notably, our work here is the first to report a magnetically saturated cavity in full GR. We observe some of the characteristics of the previously reported BMAD configuration in Newtonian MHD simulations~\cite{most_magnetically_2024, wang_bmad_2025}. However, we do not observe coherent flux bundles exit the cavity into the CBD. This is why the DH-MAD flow is a new mode of accretion. The lack of flux eruptions at the cavity-CBD wall may be driven by the strong toroidal magnetic field in the cavity which suppresses instabilities at the wall, and is likely also sensitive to the initial vertical magnetic flux in the disk. Perhaps a higher initial vertical magnetic flux in the disk will result in eruptions as in~\cite{most_magnetically_2024}. What we really need is an exploration of the CBD and magnetization parameter space in full GR, which is simply not feasible with the CPU-based GRMHD simulations we utilize. We are currently exploring this new physics with GPU-accelerated GRMHD simulations.

We also point out that `relaxing' the CBD is non-trivial. Even after $\sim 55$ binary orbits in our simulation, the cavity still undergoes morphological changes (see Figure~\ref{fig:rho_beta}). There have been many attempts to relax the CBD, usually without resolving the BH horizons or the inner cavity region ($r/M < d$, where $d$ is the binary separation)~\cite{ennoggi_merger_2025, combi_magnetized_2026}. This is an effective way to advance simulations more quickly by increasing the smallest timestep on the grid. However, with the knowledge that magnetic fields can build up in the cavity through flux eruptions and alter the accretion flow inside the cavity, it is important to resolve the horizons while relaxing the CBD~\cite{manikantan_magnetically_2025, most_magnetically_2024, wang_bmad_2025}. 

Similar long-term GRMHD simulations at a binary separation of $d/M = 30$ on an approximate, Superposed Kerr-Schild metric (SKS; \cite{combi_superposed_2021}) first relax the CBD without evolving the cavity~\cite{combi_magnetized_2026} and do not find a magnetically saturated cavity. This is likely due to the chosen initial magnetic field strengths (our initial plasma beta parameter is $\beta_{\rm ini} \sim 52$ rather than their initial density weighted average of $\beta \sim 100$). Additionally,~\cite{combi_magnetized_2026} report persistent minidisks. After our binary's MAM phase, no persistent minidisks survive in our simulation. The difference may be apparent in observations of such systems: minidisks will contribute to the highest energy thermal emission, while dynamically important magnetic fields will contribute thermal and nonthermal synchrotron emission, and produce polarized emission. Again, a systematic exploration of the CBD parameter space while evolving BH horizons (in both approximate and full GR) is necessary to understanding when magnetically saturated cavities may arise and how they effect EM observations of SMBBHs approaching merger.

Of course, our work is not without caveats. While the simulation resolution and length is at the very limit of what is achievable with CPU-based codes, it is not at the same resolution as GPU-based single BH accretion~\cite{ripperda_black_2022}. The simulation presented in this work took approximately $\sim380$ days of CPU walltime, which is not a reasonable amount for a single study. Future studies that evolve the cavity and BH horizons to self-consistently relax the CBD should utilize modern GPU-accelerated codes. Moreover, the results depend on a single binary configuration, and it is important to vary the binary and disk parameters (e.g., the net vertical magnetic flux in the initial disk) to test for the universality of the new DH-MAD accretion. This is currently impractical with CPU-based codes.
Forthcoming work of ours will allow to study the parameter space of GRMHD CBD accretion onto black holes with GPU-codes and show what key parameters determine BMAD vs DH-MAD flows.

Our work is the first to explore the magnetically saturated cavity in 3D GRMHD in full GR. The periodic horizon magnetic fluxes have significant implications for potentially periodic nonthermal flaring from SMBBHs approaching merger. The low density in the cavity and intermittent minidisks suggest that optically thin thermal and nonthermal (synchrotron) emission may dominate in the inner regions of the accretion flow. Therefore, understanding when magnetically saturated cavities may arise around SMBBHs will be crucial to identifying these systems electromagnetically and laying the groundwork for multimessenger astronomy with LISA and PTAs.

\begin{acknowledgments}
We are grateful to Elias Most, Erik Wessel, Tamara Bogdanovic, Matthew Liska, and Chris Bambic for insightful discussions during the preparation of this manuscript. We thank Mai Li for providing the thorn to compute the four-current. This work was in part supported by NASA grants 80NSSC24K0771 and 80NSSC26K0343, and NSF grant PHY-2145421 to the University of Arizona. This research is part of the Frontera computing project at the Texas Advanced Computing Center. Frontera is made possible by U.S. National Science Foundation award OAC-1818253. The work was also supported by allocation PHY190020 from the Advanced Cyberinfrastructure Coordination Ecosystem: Services \& Support (ACCESS) program, which is supported by U.S. National Science Foundation grants 2138259, 2138286, 2138307, 2137603, and 2138296~\citep{boerner_access_2023}. The Theoretical Astrophysics Program (TAP) at the University of Arizona provided resources to support this work. Generative AI (Claude Sonnet and Opus) was utilized to write code for analysis and iterate on figures. The code was verified directly by the authors. 
\end{acknowledgments}

\section*{Software and Data}

This work made use of the open-source Cactus computational framework~\cite{Goodale2003Cactus, schnetter_carpet_2016}, the Einstein Toolkit~\cite{einsteintoolkit}, \texttt{IllinoisGRMHD}~\cite{ilgrmhd}, and \texttt{kuibit}~\cite{kuibit}. This work also made use of the following software packages: \texttt{astropy} \citep{astropy:2013,astropy:2018,astropy:2022,astropy_24340642}, \texttt{Jupyter} \citep{2007CSE.....9c..21P,kluyver2016jupyter}, \texttt{matplotlib} \citep{Hunter:2007}, \texttt{numpy} \citep{numpy}, \texttt{python} \citep{python}, and \texttt{scipy} \citep{2020SciPy-NMeth,scipy_22051554}. Software citation information aggregated using \texttt{\href{https://flatironinstitute.github.io/software-citation-station/}{The Software Citation Station}} \citep{software-citation-station-paper,software-citation-station-zenodo}.

The simulation data used in this work cannot be made publicly available because the data is too large and no suitable public repository exists for hosting in this field of study. Data is available upon request.

\appendix 

\section{Longterm CBD Evolution}\label{sec:app_lump}
\begin{figure*}
    \centering
    \includegraphics[width=1\linewidth]{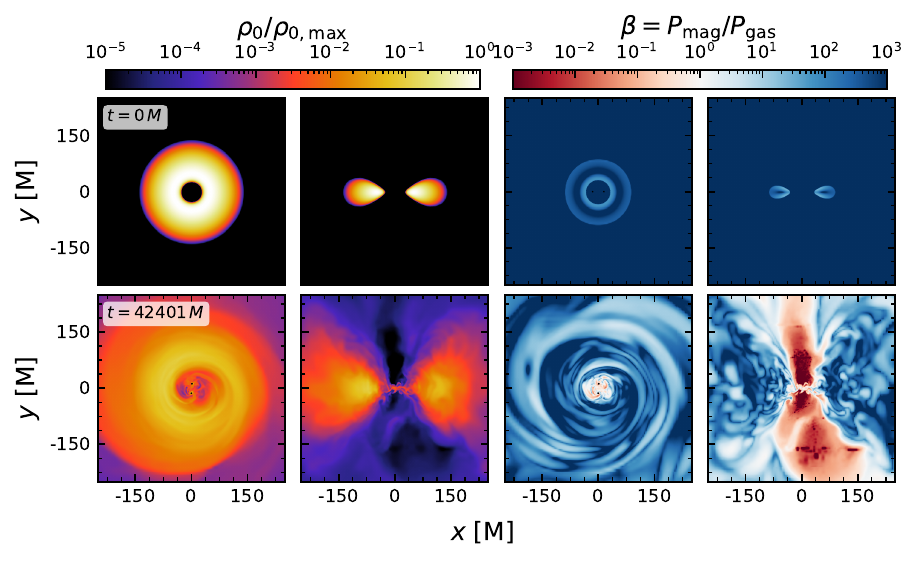}
    \caption{Initial and final snapshots of the evolved CBD. Left (right): rest-mass density (plasma beta) $xy$ and $xz$ slices at initial (top) and final (bottom) times.}
    \label{fig:full_disk}
\end{figure*}

\begin{figure}
    \centering
    \includegraphics[width=1\linewidth]{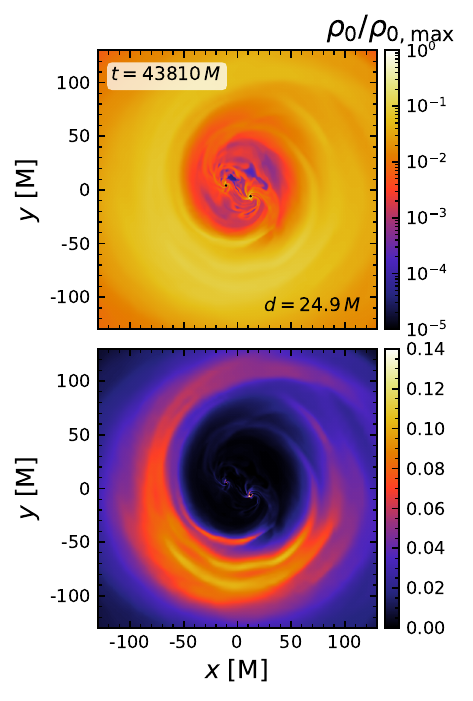}
    \caption{Rest-mass density in the $xy$ plane in log scale (top) and linear scale (bottom) to demonstrate the $m=1$ density mode (`lump') in the CBD.}
    \label{fig:lump}
\end{figure}

\begin{figure}
    \centering
    \includegraphics[width=1\linewidth]{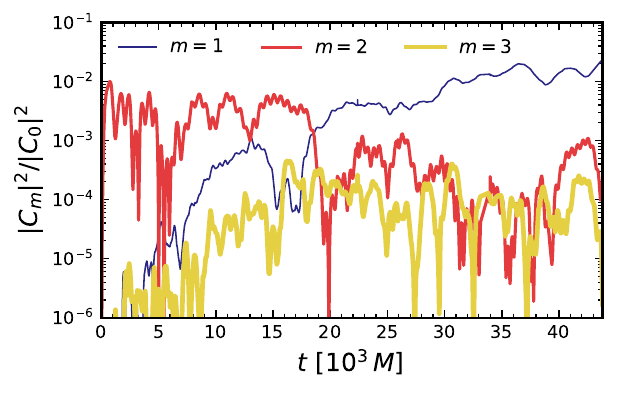}
    \caption{Power of each harmonic mode in the rest-mass density of the CBD normalized to the power of the $m=0$ mode. Initially, the $m=2$ dominates from the tidal streams perturbed by the BBH. At $t/M > 20,000$ the $m=1$ mode begins to dominate -- this is the formation of the `lump' in the CBD. The $m=3$ mode is subdominant throughout. }
    \label{fig:harmonic_power}
\end{figure}


In this section, we outline the long-term evolution of the CBD. Specifically, we describe the formation of an $m=1$ mode in the CBD and its implications for `relaxed' solutions of CBD accretion.

In Figure~\ref{fig:full_disk} we plot initial and final snapshots of the rest-mass gas density and the fluid plasma beta ($\beta \equiv P_{\rm gas}/P_{\rm mag}$) on $xy$ and $xz$ slices of the system. Initially, the CBD has an inner edge of $r/M = 27$ and outer edge $r/M \sim 140$ (top left panels). In the latest snapshot, the inner edge, which we consider roughly to be the $\rho_0/\rho_{0, \rm max} \sim 10^{-2}$ contour, ranges from $r/M \sim 30-50$ since the cavity is elliptical (bottom left panels). The minimum initial plasma beta in the CBD is $\beta \sim 26$ (top right panels). The final plasma beta in the CBD is a mix of gas pressure dominated regions ($\beta > 1$) and equipartition ($\beta \sim 1$). Above and below the cavity are regions of magnetic pressure domination $\beta < 1$ because vertical magnetic field accumulates on the BH horizons and launches dual-jets.  

Additionally, the CBD has developed an $m=1$ density mode, also known as the `lump'~\cite{macfadyen_eccentric_2008,Shilump2012ApJ...749..118S,noble_circumbinary_2012}. We highlight this feature in Figure~\ref{fig:lump}, with a logarithmic color scale (top) and a linear color scale (bottom) which demonstrates the lump feature clearly. In Fig.~\ref{fig:harmonic_power}, we plot the power of each harmonic mode (normalized by the zeroth mode) of the rest-mass density within the CBD, calculated with:

\begin{equation}
    C_m(t) \;=\; \int_{r_{\rm min}}^{r_{\rm max}} \int_0^{2\pi}
   \rho_0 \, e^{i m \phi} \; r\, dr\, d\phi,
\end{equation}
where $m$ is the azimuthal mode number, $\rho_0$ is the rest-mass density, $\phi$ is the azimuthal coordinate, $r$ is the radial coordinate, and $r_{\rm min} = 30$ and $r_{\rm max} = 200$. In the first $\sim 18$ orbits ($t/M < 19,000 \, M$) the $m=2$ dominates, because of the initial perturbation and symmetry of the BBHs. After $18$ orbits, the $m=1$ mode begins to dominate and continues to grow. The $m=3$ mode is subdominant throughout. We note that simulations of CBDs that include and evolve radiation show that this `lump' is suppressed~\cite{tiwari_radiation_2025, kim_effects_2026}.

\newpage
\bibliography{main.bib}

\end{document}